\documentclass{article}

\usepackage{graphicx}
\usepackage{amsmath,amssymb}

\usepackage[english]{babel}
\usepackage{hyperref}

\DeclareMathOperator{\tr}{tr}

\begin{document}

\thispagestyle{empty} \parskip=12pt \raggedbottom

\vspace*{1cm}
\begin{center}
  {\LARGE A specific lattice artefact in non-perturbative renormalization
    of operators.
  }

  
  \vspace{1cm} 
  Vidushi~Maillart,  Ferenc~Niedermayer \\[3mm]
  Institute for Theoretical Physics \\
  University of Bern \\
  Sidlerstrasse 5, CH-3012 Bern, Switzerland \\
  \vspace{0.5cm}
  
  \nopagebreak[4]
  
  \begin{abstract}
    We discuss a specific cut-off effect which appears 
    in applying the non-perturbative RI/MOM scheme to compute the
    renormalization constants. 
    To illustrate the problem a Dirac operator satisfying the 
    Ginsparg-Wilson relation is used, but the arguments are more general.
    We propose a simple modification of the method which
    gets rid of the corresponding discretization error.
    Applying this to full-QCD simulations done at $a=0.13\,\mathrm{fm}$
    with the Fixed Point action we find that the renormalization
    constants are strongly distorted by the artefacts discussed.
    We consider also the role of global gauge transformations,
    a freedom which still remains after the conventional
    gauge fixing procedure is applied.
  \end{abstract}

\end{center}

\eject

\section{Introduction}

The correlators of the bare fields and 
composite operators in lattice simulations are not completely physical. 
Their amplitudes depend on the 
details of the action used in the simulation and from the 
actual form of the operators involved.
To be able to compare different simulations (with different
lattice spacings, actions, operators, etc.) and also to compare
the numerical results with perturbation theory 
one needs some convention to bring these quantities
to a ``common denominator''. 
For that purpose one has to introduce renormalization 
constants, and define the renormalized fields and operators.
These will not depend any more on the microscopic details
(apart from the discretization errors) but will depend
on how the renormalization
conditions have been chosen, i.e. on the renormalization scheme.

In continuum perturbation theory one generally uses dimensional
regularization and the $\overline{\mathrm{MS}}$ scheme
to fix the convention of removing (factoring out) the divergencies.
In lattice simulations this method is not applicable, and
to be able to connect to the $\overline{\mathrm{MS}}$
scheme one has to use some condition to define the renormalized fields 
in a way which is independent of the actual regularization. 
Such scheme is the non-perturbative Rome-Southampton RI/MOM
technique \cite{Martinelli:1994ty}. 
In this scheme one fixes some renormalized Green's functions
to their free theory counterpart at some scale
$p^2=\mu^2$ where $p$ is the momentum of external particles
and $\mu$ is an appropriately chosen scale.
QCD has an extra complication due to gauge invariance:
the quark correlators are not gauge invariant, hence they average
to zero when integrated over different gauge-equivalent copies.
Therefore in the RI/MOM scheme one uses a gauge fixing condition
to make sense of the Green's functions with external
quark legs. 
This follows closely the perturbation theory where one has 
to fix the gauge anyhow. 

The RI/MOM scheme has become the standard way of fixing the 
renormalization constants, and has an extended 
literature (see 
\cite{Gockeler:1998ye,Gimenez:1998ue,Blum:2001sr,Becirevic:2004ny,Gattringer:2004iv,Aoki:2007xm}
and references therein).
In particular, there are many versions to formulate the
same renormalization condition. These coincide in the continuum
limit but contain different lattice artefacts.

We believe we found a specific artefact which can
have a large effect on the renormalization constant
and (as far as we know) all actual definitions used 
in the literature suffer from it.
Fortunately, one can cure the problem in a simple way. 
We discuss the issue for the case of the overlap 
Dirac operator where this artefact is most transparent. 
However, the problem (and the cure) is general.

The outline of the paper is the following: After stating the
conventions we describe the Landau gauge fixing, and suggest
a simplification related to the remaining global gauge 
degrees of freedom. 

The artefact in the renormalization constants is discussed in 
detail on the example of the quark field and the
composite bilinear operators.

Having modified the matching relations, 
we compare different choices for the renormalization constants 
of the quark field and of the covariant scalar density
in an actual simulation done on a $12^4$ lattice with 2+1 light flavors
with the Fixed Point (FP) action.

\section{Renormalization}

We use the following conventions for the renormalization constants.
The renormalized fermion field is related to the bare field 
as\footnote{This convention agrees with that of different
  textbooks~\cite{Itzykson:1985rh,Peskin:1995ev,Kugo:1997,DeGrand:2006}}
\begin{equation}\label{eq:psiR}
  \psi^{\mathrm{R}}= Z_q^{-1/2}\psi\,,
\end{equation}
while the renormalization factor for local operators enters 
as\footnote{{}This convention is widely followed in the general 
  and lattice literature (an exception is\cite{Peskin:1995ev})}
\begin{equation}\label{eq:OpR}
  \mathcal{O}^{\mathrm{R}} = Z_{\mathcal{O}} \mathcal{O}\,.
\end{equation}
To implement the RI/MOM scheme a gauge fixing is necessary.
One averages then the propagators over the gauge-fixed
configurations\footnote{generated according to the appropriate 
Boltzmann weight}.
The fermion propagator on a given gauge field configuration is denoted by
$S(U)$ and its gauge-field average by
\begin{equation}\label{eq:S}
  S = \langle S(U) \rangle 
\end{equation}
According to eq.~\eqref{eq:psiR} this is renormalized as
\begin{equation}\label{eq:SR}
  S^{\mathrm{R}} = 
  Z_q^{-1} S \,.
\end{equation}

\subsection{Gauge Fixing and color averaging}\label{sec:Gauge Fixing}
Eq.~\eqref{eq:S} and other gauge-field averaged 
non-gauge-invariant correlation 
functions considered below assume gauge fixing.  
We use the Landau gauge which is defined by maximizing the sum
\begin{equation}\label{eq:gauge_fixing}
  \tr \sum_{x, \mu}\left(U_\mu(x)+U^\dagger_\mu(x)\right)
\end{equation}
with respect to gauge transformations.  Here $U_\mu(x) = e^{iA_\mu(x)}\in
\mathrm{SU}(3)$ is the lattice gauge field variable.
In the continuum limit this condition is reduced to
$\partial_\mu A_\mu(x)=0$. 

A global gauge transformation $U_\mu(x)\rightarrow {^g}U_\mu(x) = g U_\mu(x)
g^\dagger$, where $g\in \mathrm{SU}(3)$ is independent of $x$, rotates the
gauge fields while keeping the value of the functional in 
eq.~\eqref{eq:gauge_fixing} unchanged. The extent of the effective 
averaging in this global degree of freedom by the available configurations 
depends not only on the number of configurations, 
but also on the actual algorithm for fixing the gauge.  
It is useful and natural to integrate over $g$ explicitly.
Consider, as an example, the propagator  
$S(x,y;U)$ on a given configuration.
Performing a global gauge transformation we get
\begin{equation}\label{eq:global_gt}
  S(x,y; U) \to  S(x,y; {^g}U)
  = g S(x,y;U) g^{\dagger} \,.
\end{equation}
When we are interested in $S(x,y)$, which is the average of the 
individual propagators over the gauge fields in Landau gauge,
as argued above, we can additionally average this quantity 
over the global gauge freedom. This ``color averaging'' amounts to
replacing the color sub-matrices (at given Dirac indices) 
by $3\times 3$ matrices which are trivial in color,
according to
\begin{equation}\label{eq:color_av}
  S(x,y) \to  \int \!\! dg \; g S(x,y) g^\dagger = 
  \mathbf{1}_\mathrm{c} \; \frac13 \tr_\mathrm{c} S(x,y) \,,
\end{equation}
where $\mathbf{1}_{\mathrm{c}}$ is the unit color matrix
and $\tr_\mathrm{c}$ denotes trace in color indices.

In the following we shall perform an additional color averaging 
whenever an averaging over gauge fields is done. 
In other words the gauge-field averaged
Green's functions will be considered to be
$4\times 4$ matrices having only Dirac indices.

Note that to perform a color averaging is mainly a matter of convenience,
it is easier to work  with $4\times 4$ than with $12\times 12$
matrices. In practice the actual color sub-matrices are already
nearly proportional to $\mathbf{1}_\mathrm{c}$.

\subsection{The renormalization conditions}\label{sec:ren_cond}
The RI/MOM technique \cite{Martinelli:1994ty} 
consists of imposing the condition that some renormalized Green's 
functions at a given scale $p^2=\mu^2$ are equal to their corresponding 
tree level values.

One would like to connect this renormalization scheme 
to other schemes by using perturbation theory and, at the
same time, control the cut-off effects on the lattice. These conditions
constrain the scale $\mu$:
\begin{equation}\label{eq:mu_cond}
  M \ll \mu \ll \frac{\pi}{a}\,,
\end{equation}
where $M$ is a typical non-Goldstone boson mass scale in QCD.
 
We can formulate an analogous condition in configuration space 
by requesting that the
renormalized Green's functions should be matched at 
distances $|x-y|$, where
\begin{equation}\label{eq:mu_condX}
  a \ll |x-y| = \frac{1}{\mu} \ll 1\,\mathrm{fm}\,.
\end{equation}

Since in full QCD applications eq.~\eqref{eq:mu_cond} 
is satisfied only marginally one needs further decisions in 
the applications in order to avoid large cut-off effects.

\subsection{A special cut-off effect rooted in lattice chiral symmetry}

\subsubsection{The quark field renormalization constant $Z'_q$
in the RI' scheme.}

We shall illustrate this cut-off effect and the way to eliminate it in the
context of the quark field renormalization factor $Z_q'$ which is
defined in the so called $\mathrm{RI}'$ 
scheme.\footnote{In the $\mathrm{RI}'$ scheme one uses $S(p)^{-1}$
while the RI scheme uses $\partial S(p)^{-1} /\partial p_\mu$ 
in the matching condition.}
In this case the Green's function considered is just 
the propagator in Fourier space given by
\begin{equation}\label{eq:Sp}
  S(p) =
  \frac{1}{V}\sum\limits_{x, y} 
  e^{-ip(x-y)} \langle \psi_x \bar\psi_y \rangle 
  = \frac{1}{V}\sum\limits_{x, y} 
  e^{-ip(x-y)} \langle S(x,y;U) \rangle 
\end{equation}
The generalization to Green's functions with a bilinear operator will 
be treated the next section.

Following the general procedure of matching with the free theory one might
request
\begin{equation}\label{eq:Zq_prime1}
  \left. S^\mathrm{R}(p) \right|_{p^2=\mu^2} = 
  Z_q^{\prime-1} \left. S(p) \right|_{p^2=\mu^2} \simeq
  \left. S(p)^{\mathit{free}}\right|_{p^2=\mu^2}\,,
\end{equation}
where $S^{\mathit{free}}$ is the inverse of the Dirac operator on the trivial
gauge configuration $U=1$ and $\mu^2$ is constrained by eq.~\eqref{eq:mu_cond}.
One might also request the same equation in coordinate space using
eq.~\eqref{eq:mu_condX}.  
Eq.~\eqref{eq:Zq_prime1} is a relation between two
$4\times 4$ matrices. If $\mu^2$ satisfies eq.~\eqref{eq:mu_cond} and the
statistical error in the simulation is small, one might expect that all
elements of this matrix equation can be (approximately) matched by a single
parameter $Z_q^\prime$.  This, however, does not hold in general.

The symmetries of the lattice action and the gauge fixing condition imply that
only the $\mathbf{1}$ and $\gamma_\mu$ Clifford algebra 
elements enter in the propagator
\begin{equation}\label{eq:Sp1g}
  S(p)  = b_0(p)\mathbf{1} + ib_\mu(p)\gamma_\mu \,.
\end{equation}
The presence of the unit $4\times 4$ matrix $\mathbf{1}$ 
is related to the $U(1)$ anomaly
and the problem of doublers~\cite{nielsenninomiya:1981}.

Consider a Dirac operator which satisfies the 
Ginsparg-Wilson (GW) relation~\cite{GW,PH_GW}
\begin{equation}\label{eq:GW}
  \gamma_5 S(U)_{xy} + S(U)_{xy}\gamma_5 =
  2 a R(U)_{xy} \gamma_5 \,,
\end{equation}
where $R(U)_{xy}$ is a local operator which is trivial in Dirac
indices. The lattice spacing $a$ is written out here explicitly
to indicate that this term vanishes in the formal continuum limit.
In particular, consider the simplest case given by Neuberger's
overlap Dirac operator \cite{Neuberger},
\begin{equation}\label{eq:overlap}
  R(U)_{xy} = \kappa \delta_{xy} \,.
\end{equation}
where $\kappa$ is a real number of order 1. 
These equations imply that the coefficient of the
unit Dirac matrix $\mathbf{1}$ in the propagator 
$D^{-1}(U)_{xy}=S(U)_{xy}$ is $a \kappa \delta_{xy}$. 
Averaging over the gauge configurations $U$ in Landau gauge
gives
\begin{equation}\label{eq:GWS}
  \gamma_5  S_{xy} + S_{xy} \gamma_5 = 
  2 a \kappa \delta_{xy}\gamma_5\,.
\end{equation}
The expectation value $S_{xy}$ depends only on $x-y$ in a periodic box. We get
for the coefficient of $\mathbf{1}$ in coordinate 
and momentum space,
\begin{equation}\label{eq:b0xy}
  b_0(x,y) = a \kappa \delta_{xy} 
\end{equation}
and
\begin{equation}\label{eq:b0p}
  b_0(p) = a \kappa \,,
\end{equation}
respectively. The matrix $S(p)^{\mathit{free}}$ has the same Dirac matrix
structure as $S(p)$
\begin{equation}\label{eq:Sp_free}
  S(p)^{\mathit{free}} = 
  b_0(p)^{\mathit{free}}\mathbf{1}
  + ib_\mu(p)^{\mathit{free}}\gamma_\mu\,,
\end{equation}
with
\begin{equation}\label{eq:b0p_free}
  b_0(p)^{\mathit{free}} = b_0(p) = a \kappa \,,\qquad \forall p\,.
\end{equation}
Matching the $b_0$ part in eq.~\eqref{eq:Zq_prime1} would give $Z'_q = 1$,
which is obviously nonsense.  
Eqs.~\eqref{eq:b0xy} and \eqref{eq:b0p} give the
explanation: $b_0$ is $\mathrm{O}(a)$, 
a pure cut-off effect in this context.
In addition, $d=|x-y|=0$ violates eq.~\eqref{eq:mu_condX}.  
Obviously, the $\mathbf{1}$ part of the propagator, whose presence 
is absolutely essential to avoid doublers and get exact chiral symmetry, 
should not be taken
(alone, or in combinations with the $\gamma_\mu$ part) in determining $Z'_q$.
For this reason the conventional matching condition 
proposed in \cite{Martinelli:1994ty} and widely used in the literature,
\begin{equation}\label{eq:old1}
  \left. Z'_q\frac{1}{4}\tr\left( 
      S(p)^{-1} i \gamma_\mu b_\mu(p)^{\mathit{free}}
    \right)\right|_{p^2=\mu^2} = 1
\end{equation}
is plagued by this artefact since the $\gamma_\mu$ part of $S(p)^{-1}$, unlike
the $\gamma_\mu$ part of $S(p)$, contains $b_0(p)$.
The same applies to an alternative definition 
\begin{equation}\label{eq:old2}
  \left. Z'_q \tr\left( S(p)^{-1} S(p)^{\mathit{free}}
    \right)\right|_{p^2=\mu^2} = 1
\end{equation}
proposed in \cite{Gattringer:2004iv}

Although we treat cut-off effects here, these distortions, as we illustrate
below, can be large if the lattice unit $a$ is not very small.

In the arguments above we discussed the case of the overlap operator.
In this case the considerations are fully transparent. 
The message is, however, general: the $\mathbf{1}$ part of the
propagator $ S(p)$ is a special cut-off effect which should not enter the
renormalization conditions.


To illustrate the point we use the results from the simulations
with 2+1 light fermions on a $12^4$ lattice, using the Fixed Point (FP) 
Dirac operator at lattice spacing $a=0.13\,\mathrm{fm}$ 
\cite{Hasenfratz:2007qe,IN_PROGRESS}
 
\begin{figure}[ht]
  \begin{center}
    \includegraphics[width=100mm]{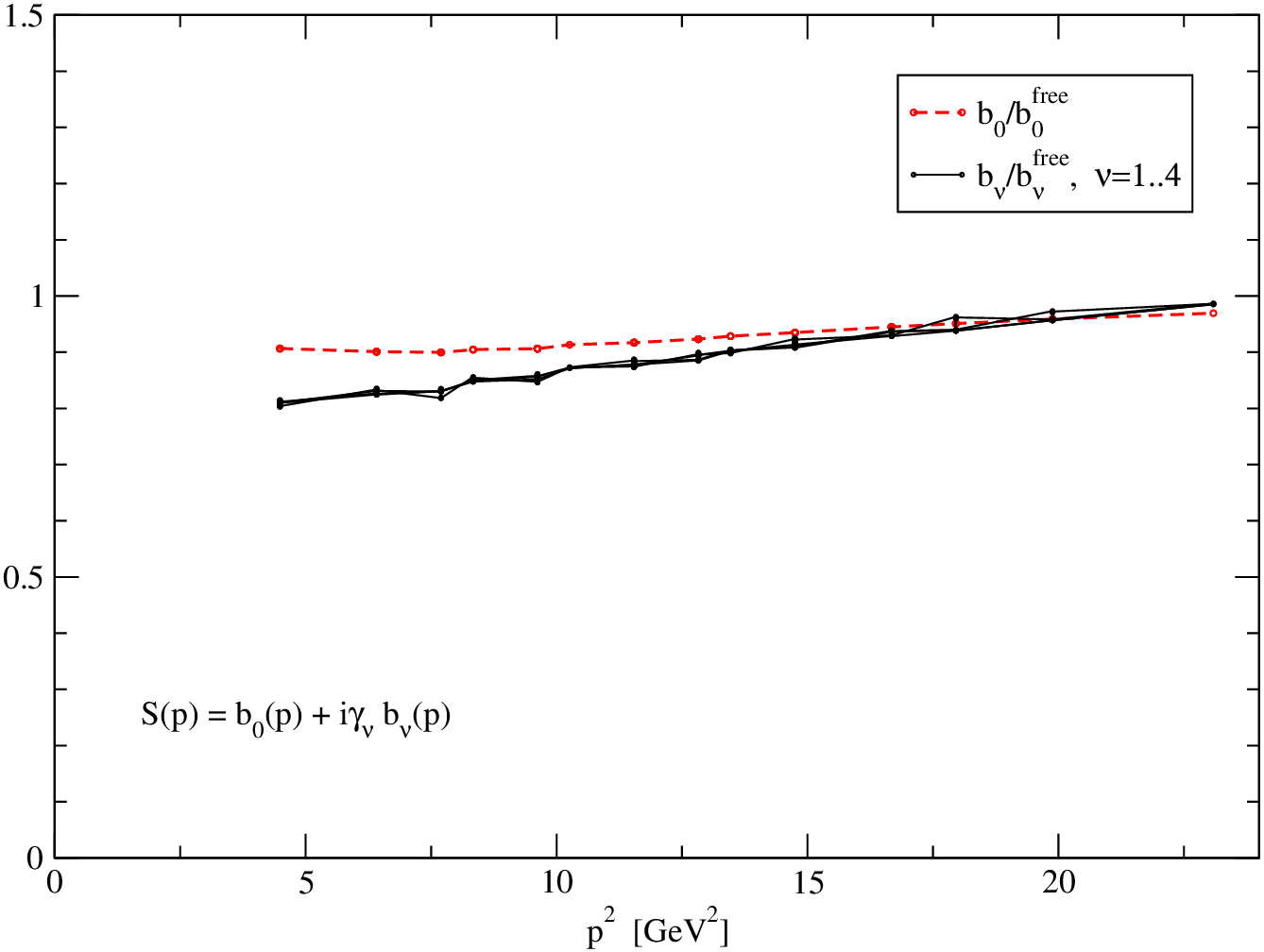}
    \caption{{} Different coefficients of the propagator 
      $S(p)=b_0(p) \mathbf{1} + i b_\nu(p) \gamma_\nu$
      divided by the corresponding coefficients of the free
      propagator.
    } \label{fig:R_Zq_prime} 
  \end{center}
\end{figure}

We plot in Fig.~\ref{fig:R_Zq_prime} the ratios
$b_0(p)/b_0(p)^{\mathit{free}}$ and $b_\mu(p)/b_\mu(p)^{\mathit{free}}$ for
$\mu=1,\ldots,4$ at different momenta $p=(p_1,p_2,p_3,p_4)$ (not averaged over
the directions of $p$). While the ratios for $\mu=1,\ldots,4$ are consistent
with each other, the $b_0$ ratio differs significantly from the others, 
in agreement with our expectations.\footnote{While our FP Dirac operator 
satisfies the GW equation to a good accuracy, the $\mathbf{1}$ 
part of the propagator is a nontrivial (nearly local) operator 
$R(x,y;U)$ hence $b_0(p)$ is expected to depend on $p$.}  
The 4-component vectors $b_\mu(p)$ and
$b_\mu(p)^{\mathit{free}}$ for a given $p$ are parallel to each other to a
{\em surprisingly high accuracy}: 
for the angle $\theta$ between them we find
$1-\cos \theta(p) \lesssim 10^{-5}$ which corresponds to an angle 
$\theta \lesssim 0.3^\circ$.

Due to this fact the matching conditions for the $\gamma_\mu$ parts are
consistent with each other.  
To be specific, we propose to average the four equations
$b_\mu(p) / Z'_q = b_\mu^{\mathit{free}}(p)$ with the weights $b_\mu(p)^{\mathit{free}}$,
i.e. to use the renormalization condition
\begin{equation} \label{eq:Zq_prime2}
  Z'_q(\mu^2)=
  \left.
    \dfrac{\sum\limits_\mu b_\mu(p) b_\mu^{\mathit{free}}(p)}{\sum\limits_\mu 
      b_\mu^{\mathit{free}}(p) b_\mu^{\mathit{free}}(p)} \right|_{p^2=\mu^2} \,.
\end{equation}
Here
\begin{equation}
  b_\mu(p)=\frac{-i}{4} \tr\left( S(p) \gamma_\mu\right) \,.
\end{equation}

It is convenient to introduce the operation of subtracting the unit 
Dirac matrix part\footnote{For later application we allow the matrix $M$ 
to have color indices as well. The color part is not affected 
by this procedure.}
\begin{equation} \label{eq:Mbar} 
  \overline{M} = M -\left(\frac{1}{4}\tr_\mathrm{D}
    M\right)\mathbf{1}\,,
\end{equation}
With this notation $\overline{S}(p)=i b_\mu(p) \gamma_\mu$ and
eq.~\eqref{eq:Zq_prime2} can be written as
\begin{equation} \label{eq:Zq_prime3} 
  Z'_q = \left. \tr\left( 
      \overline{S}(p) \overline{S}(p)^{-1}_{\mathit{free}}\right) 
    \right|_{p^2=\mu^2} \,.
\end{equation}
For consistency it is useful to check the other choice
(obtained by using $b_\mu(p)$ as weights)
\begin{equation} \label{eq:Zq_prime4} 
  \frac{1}{Z'_q} = \left. \tr\left( 
      \overline{S}(p)^{-1} \overline{S}(p)_{\mathit{free}}\right) 
    \right|_{p^2=\mu^2} \,.
\end{equation}
The two definitions should coincide if $b_\mu(p)$ is indeed
parallel to $b_\mu(p)_{\mathit{free}}$.

\begin{figure}[ht]
  \begin{center}
    \includegraphics[width=100mm]{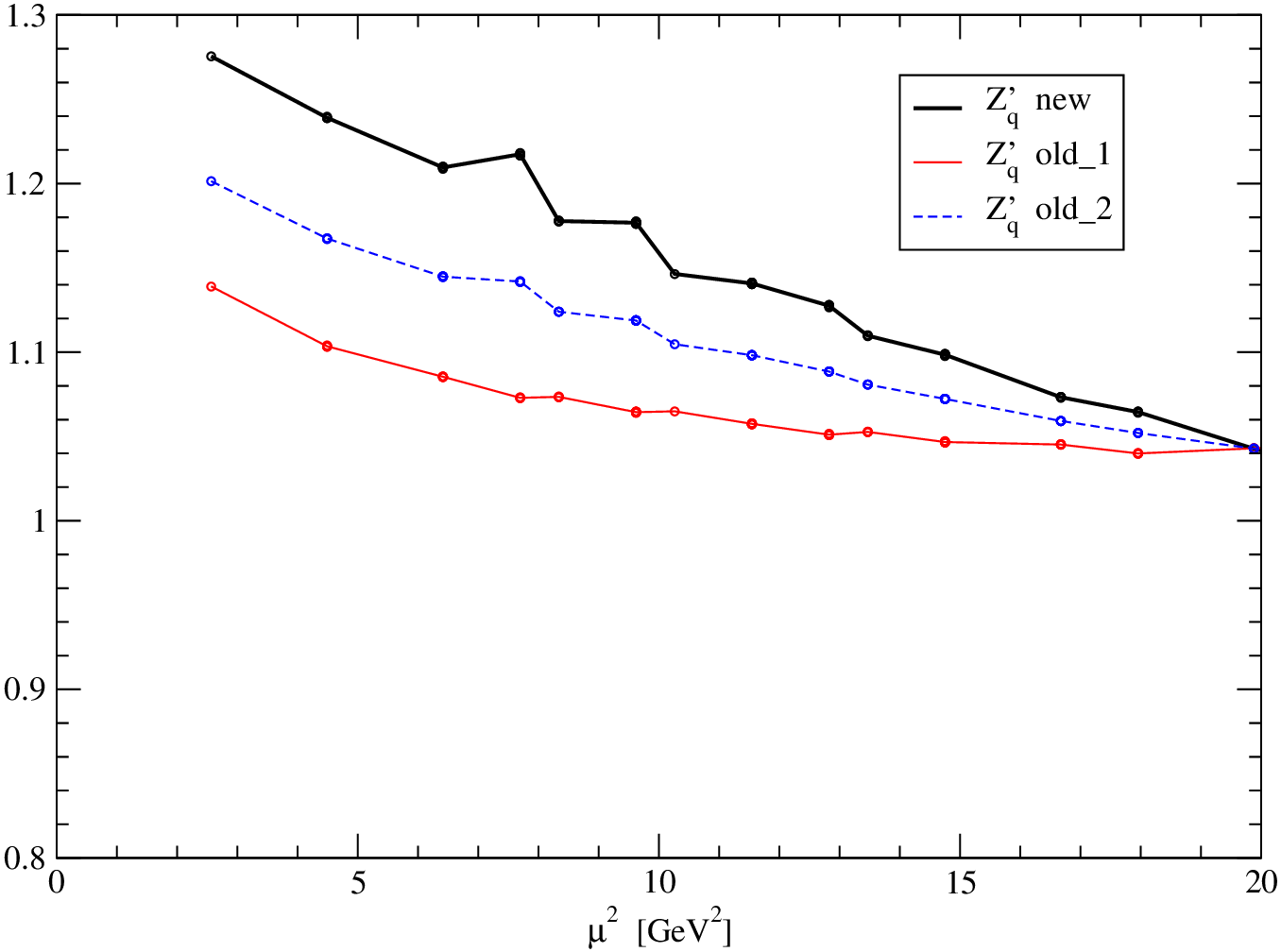}
    \caption{{} The values of $Z'_q$ calculated by different 
      methods: from eq.~\eqref{eq:Zq_prime3} (thick line), 
      and by two conventional methods, given in eq.~\eqref{eq:old1}
      (thin line) and eq.~\eqref{eq:old2} (dashed line).
  } \label{fig:Zq_prime} 
  \end{center}
\end{figure}

Fig.~\ref{fig:Zq_prime} shows the results for $Z'_q$ 
obtained by different definitions, eqs.~\eqref{eq:Zq_prime3},
\eqref{eq:old1} and \eqref{eq:old2}.

\subsubsection{Eliminating the special cut-off effects in $Z_\Gamma Z_q$}
We shall consider bilinear operators of the type\footnote{We assume for
 simplicity that $\mathcal{O}$ is non-singlet in flavor.}
\begin{equation} \label{eq:O_Gamma} \mathcal{O}_\Gamma(x) =
  \sum_{y, z} \bar{\psi}_y\Gamma(x;U)_{yz} \psi_z\,,
\end{equation}
where the kernel $\Gamma(x;U)$ is local, its value is essentially zero if the
distances $|x-y|$ or $|x-z|$ are larger than $\mathrm{O}(a)$.  
(For the naive densities and currents it has the form 
$\Gamma(x;U)_{yz}=\Gamma \delta_{xy}
\delta_{xz}$, where $\Gamma$ is the corresponding Clifford algebra element.)

The Green's function we consider is\footnote{Note that
$S(U)_{xy}$ is still a $12\times 12$ matrix; 
the color averaging is done together with averaging the whole product
over the gauge configurations.}
\begin{multline}\label{eq:GGp}
    G_\Gamma(p) = \frac{1}{V}{\sum\limits_{\substack{u,v \\ y,z}}}
    e^{-i(u-v)p} \frac{1}{V}\sum\limits_x \langle \psi_u
    \bar\psi_y\Gamma(x;U)_{yz}\psi_z\bar{\psi}_v \rangle \\
    = \frac{1}{V}{\sum\limits_{\substack{u,v \\ y,z}}}
    e^{-i(u-v)p} \frac{1}{V}\sum\limits_x \langle 
    S(U)_{uy} \Gamma(x;U)_{yz} S(U)_{zv} \rangle
\end{multline}
When one considers matching with the corresponding free field value
in coordinate space it is obvious that the distances $|u-y|$ and $|z-v|$ 
should be taken to be physical, much larger than $a$.  
As discussed previously, performing the matching in momentum space also 
contains the non-physical contact term $a \kappa \mathbf{1}$
in the case of the overlap and some similar nonphysical contribution 
for other Dirac operators. It is natural to omit these contributions
in the matching condition.

Using the notation \eqref{eq:Mbar} for subtracting the $\mathbf{1}$
part of the propagator
we define the modified Green's function
\begin{equation} \label{eq:Ghat} 
  \widehat{G}_\Gamma(p)= 
  \frac{1}{V}{\sum\limits_{\substack{u,v \\ y,z}}}
  e^{-i(u-v)p} \frac{1}{V}\sum\limits_x \langle 
  \overline{S}(U)_{uy} \Gamma(x;U)_{yz} \overline{S}(U)_{zv} \rangle \,.
\end{equation}
It is convenient to define the corresponding amputated Green's function
\begin{equation} \label{eq:Lhat} 
  \widehat{\Lambda}_\Gamma(p) =
  \overline{S}(p)^{-1}\widehat{G}_\Gamma(p) \overline{S}(p)^{-1}\,,
\end{equation}
where we use $\overline{S}(p)$ for amputation by the same argument as
discussed above. This is the quantity which we shall use in the
renormalization condition. We require that the matrix equation
\begin{equation} \label{eq:ZGamma1} 
  Z_\Gamma Z_q \left. \widehat{\Lambda}_\Gamma(p)\right|_{p^2=\mu^2}
 \simeq
 \left. \widehat{\Lambda}_\Gamma(p)^{\mathit{free}}\right|_{p^2=\mu^2}
\end{equation}
holds approximately.

Correspondingly, one can use a scalar renormalization condition 
\begin{equation}\label{eq:ZGamma}
  \left. Z_\Gamma Z_q \tr \left( \Gamma \widehat{\Lambda}_\Gamma(p)
    \right) \right|_{p^2=\mu^2} =
  \left. \tr \left( \Gamma
    \widehat{\Lambda}_\Gamma(p)^{\mathit{free}}\right) \right|_{p^2=\mu^2}
\end{equation}
for the product $Z_\Gamma Z_q$.
In other words, we replace $\Lambda_\Gamma(p)$ in the conventional lattice
definition of $Z_\Gamma Z_q$ by $\widehat{\Lambda}_\Gamma(p)$.

To illustrate the cutoff effect discussed here consider the covariant scalar
operator for a general GW Dirac operator.  (Our parametrized FP operator is an
approximation to such Dirac operator.)  In this case one has
\begin{equation} \label{eq:SRSbar} 
  S_{xy}(U) = R_{xy}(U) +
  \overline{S}_{xy}(U) \,,
\end{equation}
where $R$ is a local operator, proportional to $\mathbf{1}$ 
and appearing on the r.h.s. of the GW relation eq.~\eqref{eq:GW}, 
while $\overline{S}_{xy}(U)$ is traceless in
the Dirac indices.  The covariant scalar operator is given
essentially\cite{hasenetal:2002} by $\mathcal{O}_S = \bar{\psi} \Gamma \psi$
with $\Gamma=1/(2R)\approx 1$.  
Inserting this into eq.~\eqref{eq:GGp} one obtains four
terms, symbolically written as 
$G_S = \langle \overline{S} \Gamma \overline{S}\rangle +
\langle R \Gamma R \rangle + \text{cross terms} = \widehat{G} + G^{(1)} +
\ldots$.  For the proper matching one needs only the first term, the second
should be omitted. (The cross terms have ``wrong'' Dirac structure and do not
show up in the actual matching.)  Roughly speaking, 
$G_S \approx R/2 - b^2/(2R) \approx 1/4 - b^2$,
where $\overline{S} \sim i b_\mu \gamma_\mu$. 
Here one expects that $R(p)$ is
approximately constant while $b^2(p) \sim 1/p^2$ decreases with increasing
$p$.  

\begin{figure}[h!t]
  \begin{center}
    \includegraphics[width=100mm]{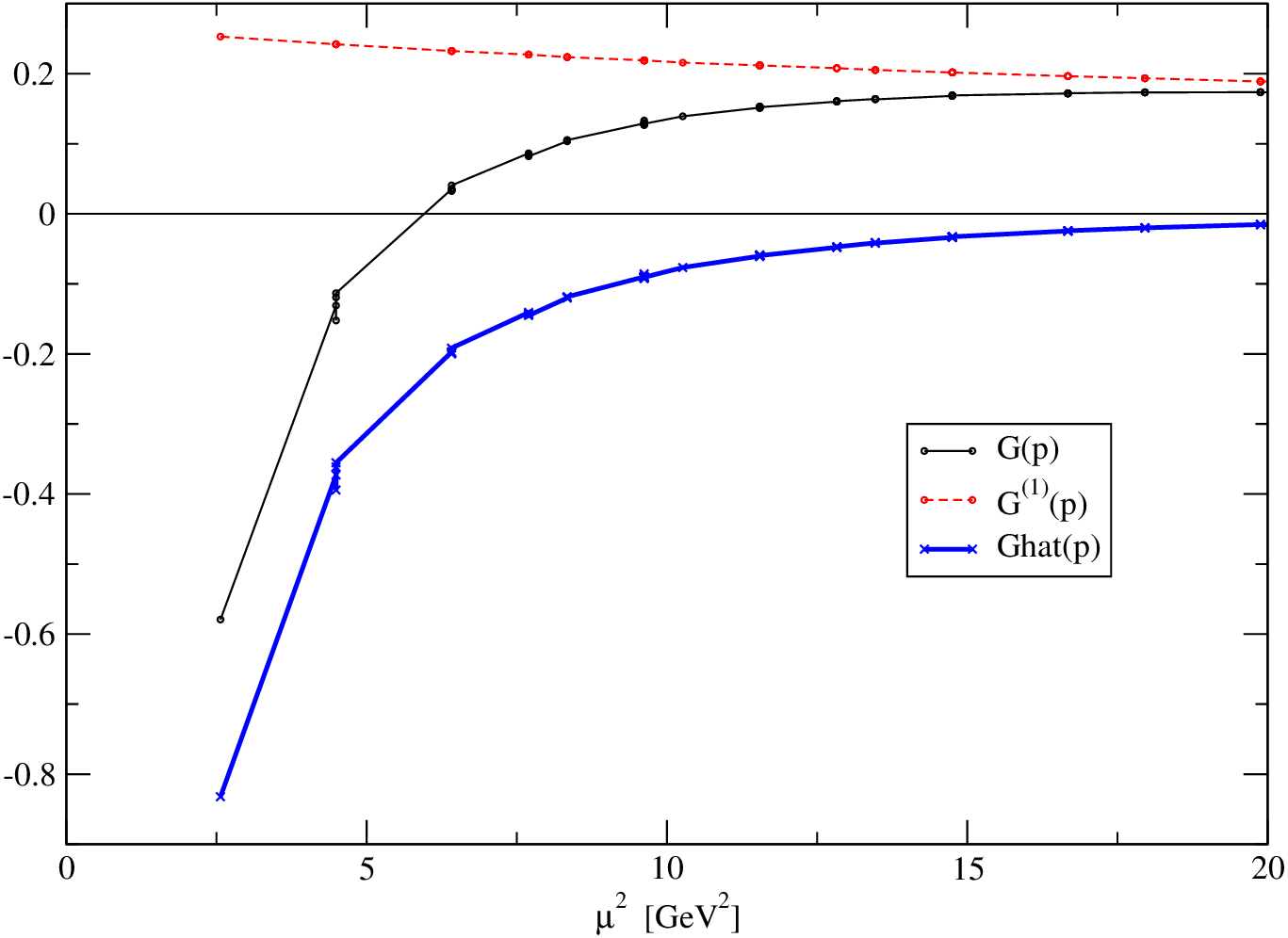}
    \caption{{} The full unamputated Green's function 
      $G(p)\sim \langle S \Gamma S \rangle$ (thin line), its
      unphysical part $G^{(1)}(p)\sim \langle R \Gamma R \rangle$
      (dashed line), and the physical part
      $\widehat{G}_\Gamma(p)\sim \langle \overline{S} 
      \Gamma \overline{S} \rangle$,
      for the scalar operator, $\Gamma=1$.
      (thick line).
  } \label{fig:G_Ghat} 
  \end{center}
\end{figure}

\begin{figure}[h!t]
  \begin{center}
    \includegraphics[width=100mm]{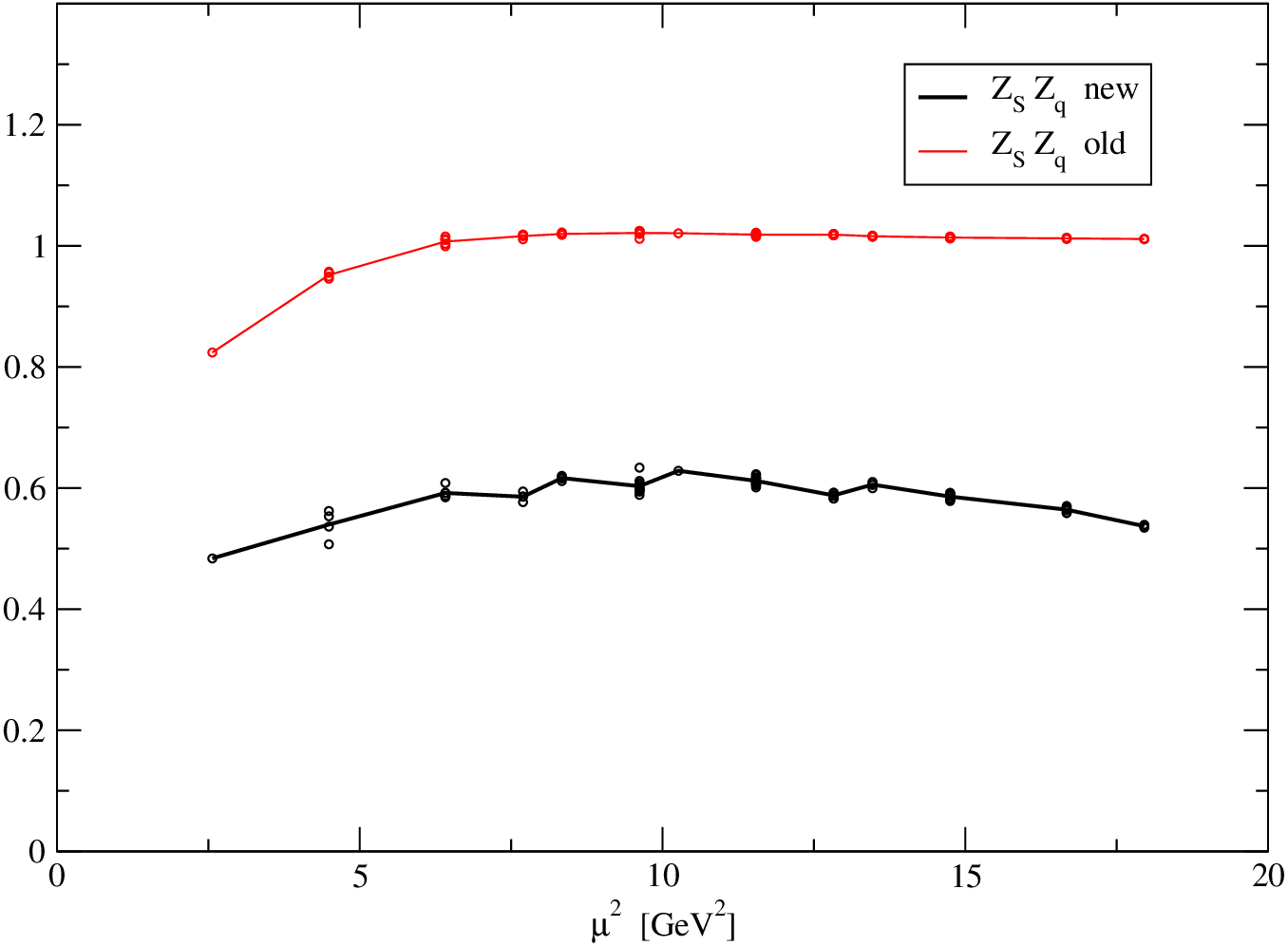}
    \caption{{} The values of $Z_S Z_q$ calculated 
      from eq.~\eqref{eq:ZGamma} (thick line), and by the 
      conventional method (thin line).
  } \label{fig:ZZS} 
  \end{center}
\end{figure}

In Fig.~\ref{fig:G_Ghat} we plot the full $G_S$ 
and the terms $\widehat{G}_S$ and $G^{(1)}_S$ using our data.  The figure shows that
the nonphysical part $G^{(1)}_S$ dominates for $p^2 > 5 \,\mathrm{GeV}^2$
hence taking the full quantity $G_S$ instead of $\widehat{G}_S$ results 
in a very strong distortion of the true matching ratio.
Fig.~\ref{fig:ZZS} illustrates how 
the proposed modifications affect the value $Z_\Gamma Z_q$.

\subsubsection{The quark field renormalization constant $Z_q$
in the RI scheme.}

Consider now the special case when the bilinear operator
$\mathcal{O}_\Gamma(x)$ is the conserved current (related to the flavor
conservation):
\begin{equation}\label{eq:Vc_mu}
  V^c_\mu(x) = \bar{\psi}_y K_\mu(x;U)_{yz}\psi_z\,.
\end{equation}
Given the Dirac operator $D(U)$, there is a simple procedure to find
$K_\mu$~\cite{Kikukawa,hasenetal:2002}.  The kernel
satisfies the Ward identity in coordinate space
\begin{equation}\label{eq:KmuxU}
  \sum\limits_\mu \partial^{*}_\mu K_\mu(x;U)_{yz} = 
  \left(\delta_{xy} - \delta_{xz}\right)D(U)_{yz}\,,
\end{equation}
where $\partial^{*}_\mu$ is the backward lattice derivative in $x$.  Starting
from eq.~\eqref{eq:KmuxU} it is not difficult to get 
the Ward identity for the corresponding Green's function in momentum space
\begin{equation}\label{eq:GWI}
  G_\mu^{V^c}(p) = -i\frac{\partial}{\partial p_\mu} S(p)\,,
\end{equation}
where $G_\mu^{V_c}(p)$ is defined in eq.~\eqref{eq:GGp} with $\Gamma(x)_{yz}\to
K_\mu(x;U)_{yz}$. For the amputated Green's function we get
\begin{equation}\label{eq:LWI}
  \Lambda_\mu^{V_c}(p)=  S(p)^{-1} G_\mu^{V_c}(p)
  S(p)^{-1} 
  =i\frac{\partial}{\partial p_\mu} S(p)^{-1}\,.
\end{equation}
The quantity on the r.h.s., $\partial S(p)^{-1}/\partial p_\mu$ is used to fix
the renormalization factor of the quark field.  
The corresponding condition defines a
scheme which is different from that given by  eqs.~\eqref{eq:Zq_prime1} and
\eqref{eq:Zq_prime2}. The quark field renormalization factor in this scheme is
denoted by $Z_q$.  The Ward identity shows that in this scheme the conserved
current does not renormalize, i.e. $Z_{V^c}=1$.  
(The advantage of using the vertex function of the conserved current 
lies in the fact that it avoids approximating the derivative
over the momentum by discrete derivatives available on a finite lattice.)

Note that the conventional definition of $Z_q$ through eq.~\eqref{eq:LWI}
\begin{equation}\label{}
  \left. Z_q \tr \left( \gamma_\mu
      \frac{\partial S(p)^{-1}}{\partial p_\mu}\right) \right|_{p^2=\mu^2} 
  =
  \left. \tr \left( \gamma_\mu
    \frac{\partial S(p)^{-1}_{\mathit{free}}}{\partial p_\mu}
    \right)
  \right|_{p^2=\mu^2}\,,
\end{equation}
also suffers from the same lattice artefact. Instead of this
condition  one should take
\begin{equation}\label{eq:Zq1}
  Z_q \left. 
    \frac{\partial \overline{S}(p)^{-1}}{\partial p_\mu}\right|_{p^2=\mu^2} 
  \simeq
  \left. \frac{\partial \overline{S}(p)^{-1}_{\mathit{free}}}%
    {\partial p_\mu}\right|_{p^2=\mu^2} \,,
\end{equation}
or equivalently
\begin{equation}\label{eq:Zq2}
  Z_q \left. \overline{S}(p)^{-1} \overline{G}_\mu^{V_c}(p)
    \overline{S}(p)^{-1} \right|_{p^2=\mu^2} 
  \simeq
  \left. \overline{S}(p)^{-1}_{\mathit{free}} 
    \overline{G}_\mu^{V_c}(p)_{\mathit{free}}
    \overline{S}(p)^{-1}_{\mathit{free}} \right|_{p^2=\mu^2} \,.
\end{equation}
To have a scalar equation it is convenient to take out the $\gamma_\mu$
part and sum over $\mu$
\begin{equation}\label{eq:Zq3}
  Z_q \left. \tr \left( \gamma_\mu \overline{S}(p)^{-1} 
      \overline{G}_\mu^{V_c}(p)
    \overline{S}(p)^{-1} \right) \right|_{p^2=\mu^2} =
  \left. \tr \left( \gamma_\mu \overline{S}(p)^{-1}_{\mathit{free}} 
    \overline{G}_\mu^{V_c}(p)_{\mathit{free}}
    \overline{S}(p)^{-1}_{\mathit{free}}\right) \right|_{p^2=\mu^2} \,.
\end{equation}

Note that here we have $\overline{G}_\mu^{V_c}$ which at first sight differs
from $\widehat{G}_\mu^{V_c}$ defined in eq.~\eqref{eq:Ghat}.  
However, due to the Ward identity the Green's function for 
the conserved current $G_\mu(p)$ is effectively linear in the propagator 
$S(p)$ and the two quantities coincide. 
For general operators, however, one should use the quantities 
$\widehat{G}_\Gamma$ or $\widehat{\Lambda}_\Gamma$.

\begin{figure}[h!t]
  \begin{center}
    \includegraphics[width=100mm]{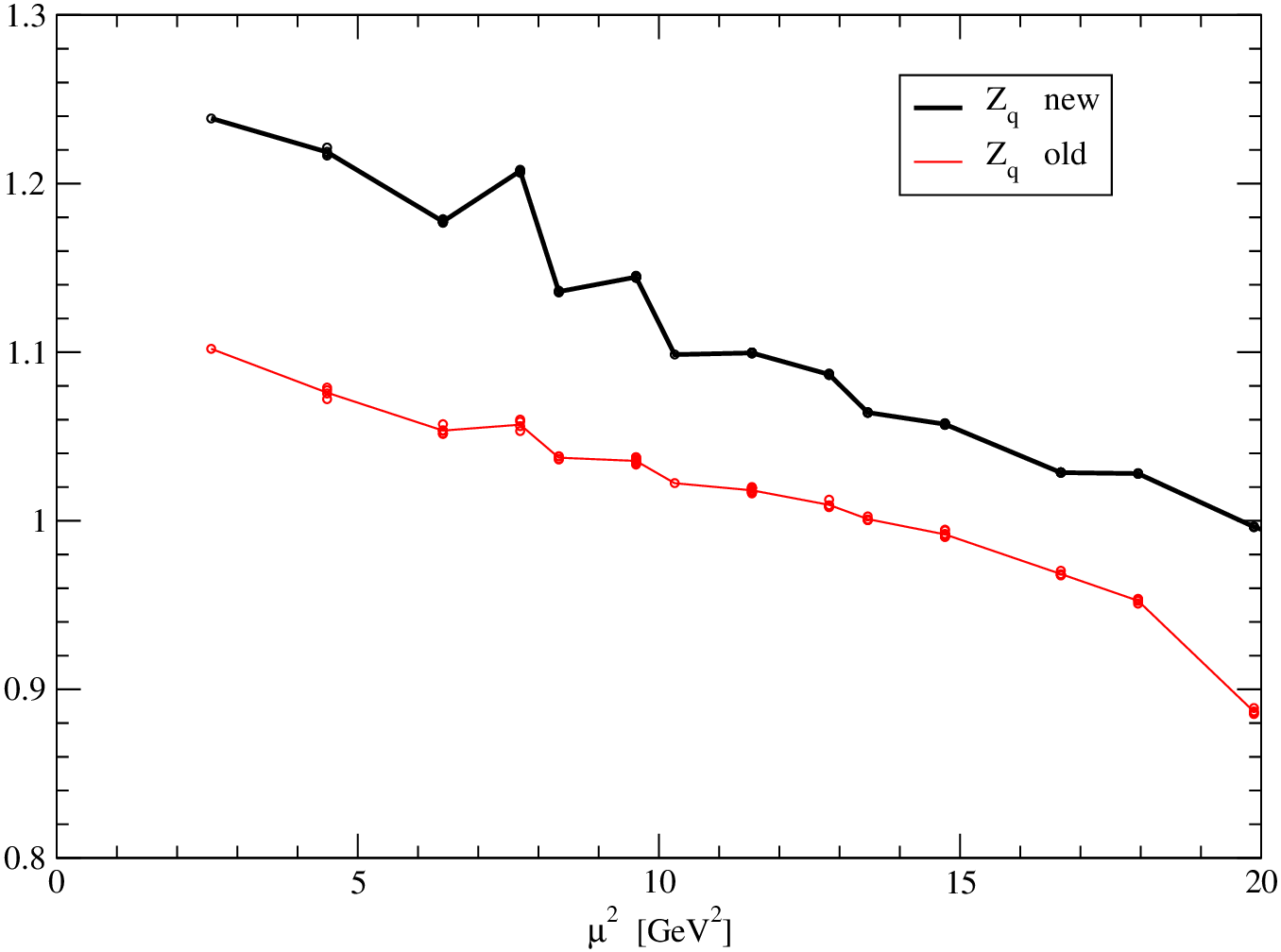}
    \caption{{} The values of $Z_q$ calculated by different 
      methods: from eq.~\eqref{eq:Zq2} (thick line), and by the 
      conventional method (thin line).
  } \label{fig:Zq} 
  \end{center}
\end{figure}

\section{Summary}

We suggested to color average the gauge-field averaged
propagators and Green's functions used in the RI/MOM scheme, 
reducing them to $4\times 4$ matrices having only Dirac indices.
This is, however, just a matter of convenience.

Further, we argued that the presence of the unphysical, short range part
of the quark propagator proportional to the unit $4\times 4$
matrix (in Dirac indices) is responsible for a considerable
O($a$) artefact in the renormalization constants of the quark field
and the composite operators. We illustrated that by omitting this
contribution one can satisfy the matching condition in the matrix sense
to a remarkable precision, which is spoiled otherwise by the non-physical
contribution.

The effect discussed here is quite large for the lattice simulations
with the Fixed Point fermions at a lattice spacing $a=0.13\,\mathrm{fm}$.
However, the discretization error is expected to be less pronounced 
in simulations at smaller lattice spacing. 
It can also be smaller for the Wilson action 
where the coefficient of the $\mathbf{1}$ in the propagator can 
decay faster in $p$. These questions need further study.

{\bf Acknowledgement} We thank Peter Hasenfratz for enlightening
discussions on the problem of these artefacts, 
Anna Hasenfratz for sharing her experience with the
RI/MOM scheme and Dieter Hierl for helping us with the numerical data.

\clearpage


\begin{thebibliography}{1}

\bibitem{Martinelli:1994ty}
  G.~Martinelli, C.~Pittori, C.~T.~Sachrajda, M.~Testa and A.~Vladikas,
  Nucl.\ Phys.\  B {\bf 445} (1995) 81
  [arXiv:hep-lat/9411010].



\bibitem{Gockeler:1998ye}
  M.~Gockeler {\it et al.},
  Nucl.\ Phys.\  B {\bf 544} (1999) 699
  [arXiv:hep-lat/9807044].


\bibitem{Gimenez:1998ue}
  V.~Gimenez, L.~Giusti, F.~Rapuano and M.~Talevi,
  Nucl.\ Phys.\  B {\bf 531}, 429 (1998)
  [arXiv:hep-lat/9806006].

\bibitem{Blum:2001sr}
  T.~Blum {\it et al.},
  Phys.\ Rev.\  D {\bf 66}, 014504 (2002)
  [arXiv:hep-lat/0102005].

\bibitem{Becirevic:2004ny}
  D.~Becirevic, V.~Gimenez, V.~Lubicz, G.~Martinelli, M.~Papinutto and J.~Reyes,
  JHEP {\bf 0408}, 022 (2004)
  [arXiv:hep-lat/0401033].

\bibitem{Gattringer:2004iv}
  C.~Gattringer, M.~Gockeler, P.~Huber and C.~B.~Lang,
  Nucl.\ Phys.\  B {\bf 694}, 170 (2004)
  [arXiv:hep-lat/0404006].

\bibitem{Aoki:2007xm}
  Y.~Aoki {\it et al.},
  arXiv:0712.1061 [hep-lat].


\bibitem{Itzykson:1985rh} C. Itzykson and J.~B. Zuber, \emph{Quantum Field
    Theory}, McGraw-Hill (1985), \ p.\ 705.

\bibitem{Peskin:1995ev} M.~E. Peskin and D.~V. Schroeder, \emph{An
    Introduction to Quantum Field Theory}, Perseus Books (1995), \ p.\ 842.

\bibitem{Kugo:1997} T. Kugo, \emph{Eichtheorie}, Springer (1997), \ p.\ 522.


\bibitem{DeGrand:2006} 
T.~DeGrand and C.~DeTar, 
\emph{Lattice Methods for Quantum Chromodynamics}, 
World Scientific Publishing Company (2006), \ p.\ 364.


\bibitem{nielsenninomiya:1981} N.~B. Nielsen and M. Ninomiya, Phys. Lett.
  {\bfseries B105} (1981), 219; Nucl. Phys. {\bfseries B185} (1981), 20.



\bibitem{GW} 
  P.~H.~Ginsparg and K.~G.~Wilson,
  Phys.\ Rev.\ D {\bf 25} (1982) 2649.


\bibitem{PH_GW}
  P.~Hasenfratz,
  Nucl.\ Phys.\  B {\bf 525}, 401 (1998)
  [arXiv:hep-lat/9802007].


\bibitem{Neuberger}
  H.~Neuberger,
  Phys.\ Lett.\ B {\bf 417} (1998) 141
  [arXiv:hep-lat/9707022].




\bibitem{Hasenfratz:2007qe}
  P.~Hasenfratz, D.~Hierl, V.~Maillart, F.~Niedermayer, A.~Schafer,
  C.~Weiermann and M.~Weingart, 
  PoS {\bf LAT2007}, 077 (2007)
  [arXiv:0710.0551 [hep-lat]].

\bibitem{IN_PROGRESS} P.~Hasenfratz, D.~Hierl, V.~Maillart, F.~Niedermayer,
  A.~Schafer, C.~Weiermann and M.~Weingart, in progress



\bibitem{hasenetal:2002}
  P.~Hasenfratz, S.~Hauswirth, T.~Jorg, F.~Niedermayer and K.~Holland,
  Nucl.\ Phys.\  B {\bf 643} (2002) 280
  [arXiv:hep-lat/0205010].


\bibitem{Kikukawa}
  Y.~Kikukawa and A.~Yamada,
  [arXiv:hep-lat/9810024],


\end{thebibliography}
\end{document}